\documentclass[twocolumn,prl,aps]{revtex4-1}
\usepackage{amsmath}
\usepackage{graphicx}
\usepackage{sistyle}
\usepackage{color}

\begin{document}

\title{Quantum random number generation on a mobile phone}
\author{Bruno Sanguinetti}
\email{Bruno.Sanguinetti@unige.ch}
\address{Group of Applied Physics, University of Geneva, Switzerland}
\author{Anthony Martin}
\address{Group of Applied Physics, University of Geneva, Switzerland}
\author{Hugo Zbinden}
\address{Group of Applied Physics, University of Geneva, Switzerland}
\author{Nicolas Gisin}
\address{Group of Applied Physics, University of Geneva, Switzerland}

\begin{abstract}
Quantum random number generators (QRNGs) can significantly improve the security of cryptographic protocols, by ensuring that generated keys cannot be predicted. However, the cost, size, and power requirements of current QRNGs has prevented them from becoming widespread. In the meantime, the quality of the cameras integrated in mobile telephones has improved significantly, so that now they are sensitive to light at the few-photon level. We demonstrate how these can be used to generate random numbers of a quantum origin.
\end{abstract}
\maketitle

\section{Introduction}
The security of cryptographic protocols, both classical and quantum, relies on the generation of high quality random numbers. For example,  classical asymmetric key protocols such as DSA~\cite{Kravitz1993}, RSA \cite{Rivest1978,Adleman1983} and Diffie-Hellman~\cite{Diffie1980}, use random numbers, tested for primality, to generate their keys. Another example is the unconditionally secure one-time pad protocol, which needs a string of perfectly random numbers of a length equal to that of the data to be encrypted. The main limitation of this protocol is the requirement for key exchange. Quantum key distribution offer a way to generate two secure keys at distant locations, but its implementation also requires a vast quantity of random numbers~\cite{Gisin2002}.

Famously, Kerckhoffs's principle~\cite{Kerckhoffs1883} states that the security of a cypher must reside entirely in the key. It is therefore of particular importance that the key is secure, which in practice requires it to be chosen at random. In the past, weaknesses in random number generation~\cite{Lenstra2012} have resulted in the breaking of a number of systems and protocols, such as operating system security~\cite{Dorrendorf2009}, communication protocols~\cite{Bello2008}, digital rights management~\cite{Bushing2010} and financial systems~\cite{Chirgwin2013}.

High quality random numbers are hard to produce, in particular they cannot be generated by a deterministic algorithm such as a computer program. To ensure the randomness, and importantly, the uniqueness of the generated bit string, a physical random number generator is required~\cite{Vincent1970,Saitoh2005}. Of particular interest are quantum random number generators (QRNGs)\cite{Rarity1994}, which by their nature produce a string which cannot be predicted, even if an attacker has complete information on the device. QRNGs have typically been based on specialised hardware, such as single photon sources and detectors~\cite{Stefanov2000,Dultz2002,Wei2009} or homodyne detection~\cite{Gabriel2010,Shen2010}.  Image sensors have been used to generate random numbers of classical origin by extracting information from a moving scene, e.g. a lava lamp, or using sensor readout noise~\cite{Mende1998} but their performance both in terms of randomness and throughput has been low. Here we show how random numbers of a quantum origin can be extracted from an illuminated image sensor. Nowadays, cameras are integrated in many common devices such as cell phones, tablets and laptops.

In the first part of this paper we describe the concept of our system, including its various entropy sources and how the entropy of quantum origin can be extracted. In the second part, we characterise two different cameras for random number generation. Finally we present our results and test the generated random numbers.

\section{Concept}

Most light sources emit photons at random times. Thus, it is impossible to perfectly define the number of photons emitted per unit time. This quantum effect is usually called ``quantum noise'' or ``shot noise'' and has been shown to be a property of the light field rather than the detector~\cite{Brida2010}. Only some particular light sources, namely amplitude-squeezed light~\cite{Walls1983}, can overcome this fundamental noise.
Beside these very specific sources, the number of photons emitted per unit of time is governed by a Poisson distribution. For a mean number of photons $\bar n$, we obtain a standard deviation of $\sqrt{\bar n}$. We can exploit this quantum effect to realise a QRNG by using a detector capable of resolving this distribution.

\begin{figure}
  \centering
  \includegraphics[width=\linewidth]{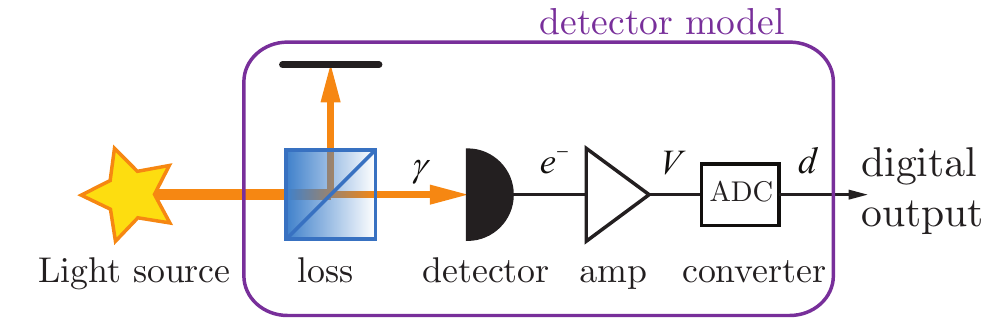}\\
  \caption{A detector, or indeed each pixel of an image sensor, can be modelled as having 100\% efficiency, but being preceded by a lossy element (beamsplitter) with transmission $\eta$. For each absorbed photon, the detector generates an electron. This charge is then converted into a voltage and amplified, before being digitised and sent to further processing, i.e. a randomness extraction stage. }\label{fig_concept}
\end{figure}

\begin{figure}[htbp]
\begin{center}
\includegraphics[width=0.9\columnwidth]{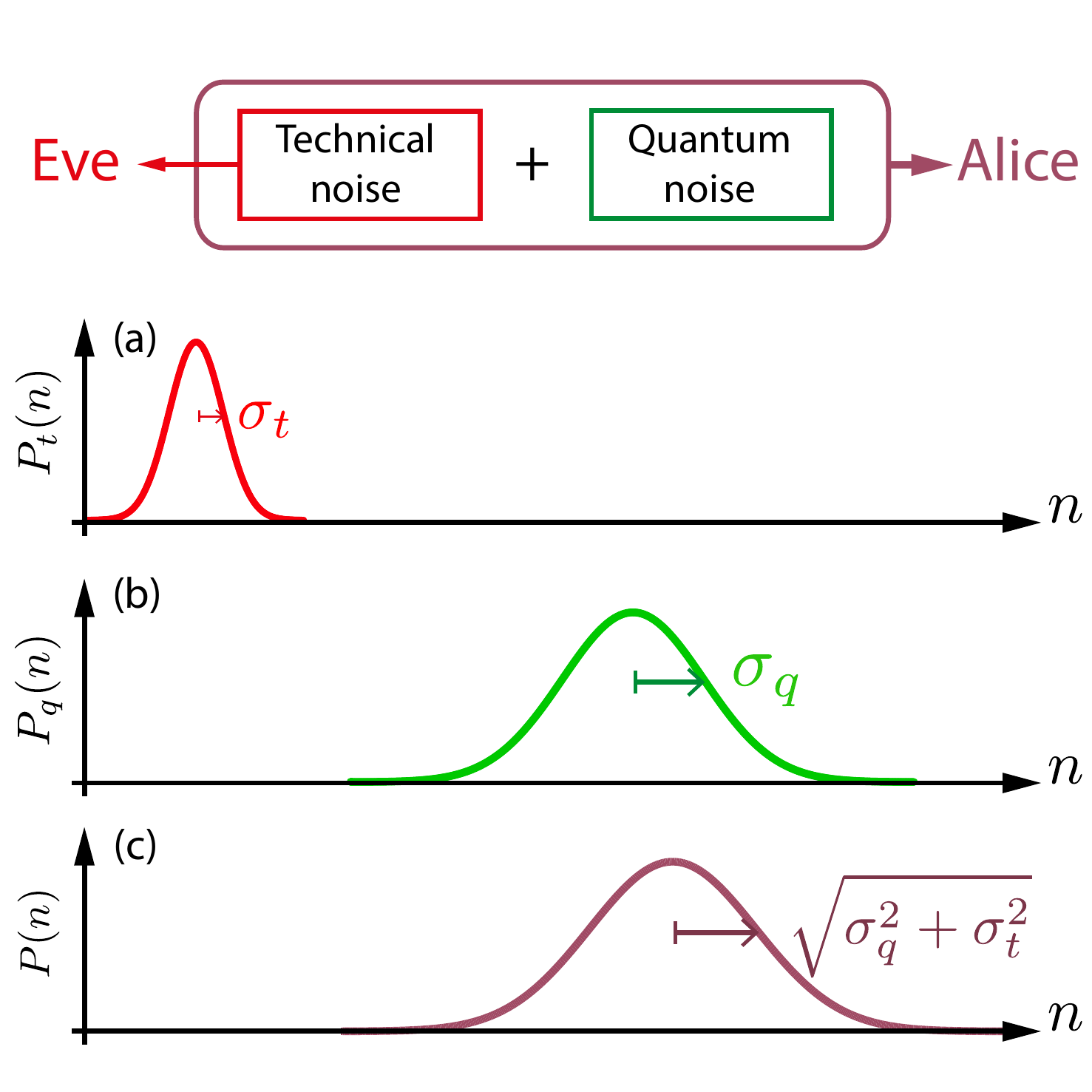}
\caption{Working principle and assumptions: (a) we measure a number $n$ of photoelectrons on a image sensor's pixel with a probability $P(n)$. Assuming that the detector is operating in a linear regime, this measured distribution will be the combination of quantum uncertainty (b) and technical noise (c). From a single shot measurement we cannot distinguish these two noise components, however we assume that to our adversary the technical noise is fully deterministic.}
\label{fig_quantum_classical_noise}
\end{center}
\end{figure}

As shown in \figurename{~\ref{fig_concept}}, a detector can be modelled as lossy channel with a transmission probability $\eta$ follow by a photon-to-electron converter with unit efficiency. In this model, $\eta$ contains all the losses due to the optical elements and the detector's quantum efficiency. An analogue-to-digital converter (ADC) encodes the electron numbers into digital values.
We can define an electron-to-digital conversion factor $\zeta$. If $\zeta\geq1$ for each possible number of electrons there is at last one unique corresponding digital code.
Under these conditions we access the shot noise statistics of the light and can use this to generate quantum random numbers.
To complete the model of the detector, noise needs to be added. This noise has different origins \emph{e.g.} thermal noise, leakage current or readout noise. Generally, this noise follows a normal distribution and adds linearly to the signal, as show in \figurename{~\ref{fig_quantum_classical_noise}}.

At the output of the detector, we obtain a random variable $X=X_q + X_t$, where $X_t$ and $X_q$ are independent random variables taken from the technical noise distribution $\mathcal{D}_t$ and the quantum uncertainty distribution $\mathcal{D}_q$, respectively. We assume that the technical noise is completely known to an adversary (Eve). We can thus rely only on the quantum entropy generated.

The amount of quantum entropy will correspond to the entropy of a Poisson distribution with a mean equal to the average number of photons absorbed $\bar n$, which expressed in bits is:

\begin{equation}
H(X_q)=\frac{\bar{n}}{\ln(2)}[1 - \ln(\bar{n})] + \frac{e^{-\bar{n}}}{\ln(2)}\sum_{m=0}^\infty \frac{\bar{n}^m\ln(m!)}{m!}
\label{eq:entropy}
\end{equation}
for large values of $\bar{n}$ this expression can be approximated to:
\begin{equation}
H(X_q)\approx\frac{\ln(2\pi e\bar{n})}{2\ln{2}}.
\end{equation}
To collect this entropy entirely, the detector must have $\zeta\geq 1$.
The measured value $X$ is encoded over $b$ bits. The entropy $H(X_q)$ of quantum origin per bit of output will be on average $H(X_q)/b<1$.
To obtain a string of perfectly random bits, \textit{i.e.} with unit quantum entropy per bit, an extractor is required.

As detailed in Ref.~\cite{Troyer2012}, an extractor computes a number $k$ of high-entropy output bits $y_j$ from a number $l>k$ of lower-entropy input bits $r_i$.
This can be done by performing a vector-matrix multiplication between the vector formed by the raw bit values $r_i$ and a random $l\times k$ matrix $M$ (performed modulo 2):
\begin{equation}
y_j = \sum^l_{i=1}M_{ji}r_i.
\label{eq:extractor}
\end{equation}
Note that although the element of $M$ are randomly distributed, $M$ is a pre-generated constant.
For raw input bits with entropy $s$ per bit, the probability that the output vector $y_j$ deviates from a perfectly random bit string is bounded by:
\begin{equation}
\epsilon=2^{-(s\,l-k)/2}.
\end{equation}


\section{Experiment}
Detectors able to resolve shot noise have traditionally been complicated and bulky, e.g. homodyne detectors.
In recent years, however, image sensors such as the ones found in digital cameras and smartphones have improved immensely. Their readout noise is of the order of a few electrons and their quantum efficiencies can achieve 80\%. Besides their ability to resolve quantum noise with high accuracy, image sensors are intrinsically parallel and offer high data rates.
Here we generate quantum random numbers both with a commercial astronomy monochrome CCD camera (ATIK 383L), and a CMOS sensor in a mobile phone (Nokia N9), a colour camera, from which we use only the green pixels for the purpose of this article.

\begin{figure}
  \centering
  \includegraphics[width=\columnwidth]{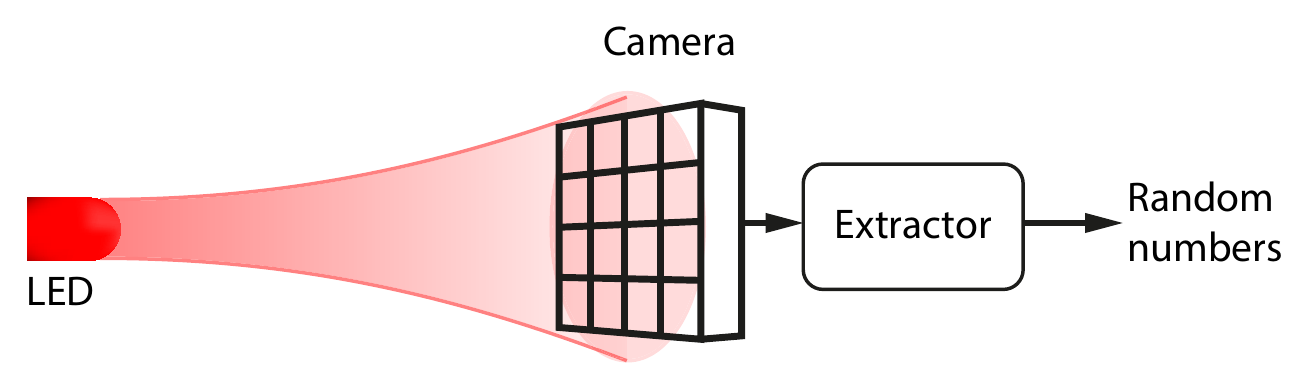}\\
  \caption{Random number generator setup: a camera is fully and homogeneously illuminated by a LED. The raw binary representation of pixel values are concatenated and passed through a randomness extractor. This extractor outputs quantum random numbers. }\label{fig:camera_experiment}
\end{figure}

The experimental setup for the random number generator is shown in Fig.~\ref{fig:camera_experiment}: a camera is illuminated by a LED, the raw pixel data is passed through an extractor the output of which are random numbers ready to be used.

First however, we check that the cameras comply with the manufacturer's specification and that the operating conditions are appropriate for the generation of quantum random numbers. In particular, we are interested in verifying that the photon number distribution does not exceed the region where the camera is linear, and that there are enough digital codes to represent each possible number of absorbed photons, i.e. $\zeta\geq 1$.
\subsection{Characterization}
To characterise the two cameras, we use a well controlled light source based on a light emitting diode (LED), as shown in Fig.~{~\ref{fig:camera_experiment}}.

As shown in Fig.~\ref{fig_concept}, a number of photons $n$ is absorbed by the image sensor and converted into an equal number of electrons. This charge is in turn converted into a voltage by an amplifier, and finally digitised. The amplifier gain (which in the sensors used corresponds to ``ISO'' setting) is set such that each additional input electron will result in an output voltage increase sufficient to be resolved by the ADC. This means that each electron increase the digital output code $c$ by at least 1. We check that this is the case by illuminating the cameras with a known amount of light. Using the nominal quantum efficiency of the cameras we can infer $\bar n$, and observe $\zeta=c/e$ to be 2.3 codes/electron for the ATIK camera, and 1.9 codes/electron for the Nokia camera, as expected from the devices' specifications.

\begin{figure}[htbp]
\begin{center}
\includegraphics[width=0.98\columnwidth]{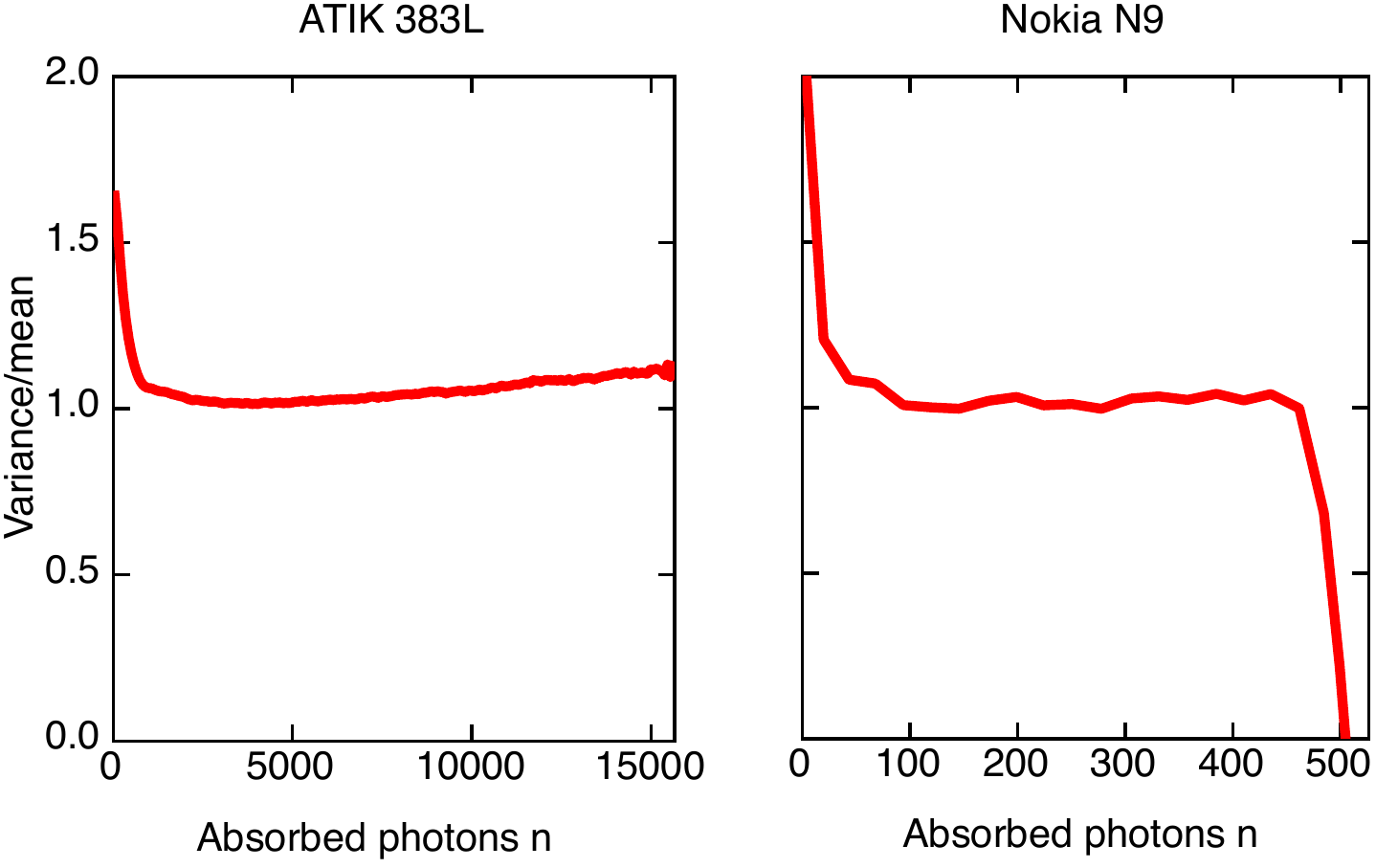}
\caption{Fano factor (Variance/mean) of the devices emptied in this experiment. We operate in the region where the Fano factor is 1 and the detector is most linear.}
\label{fig_Fano}
\end{center}
\end{figure}

To evaluate the linearity of the camera sensors, we measure the Fano factor given by $F=\frac{{\rm Var(c)}}{\zeta c}$.
In Fig.~\ref{fig_Fano} we plot the $F$ for various illuminating intensities of our light sensors. Both detectors have a large range of intensities where the Fano factor is constante with a value close to 1. In this range the statistics are dominated by the quantum uncertainty (shot noise). At strong illuminations, saturation occurs, for the Nokia N9 this happens at intensities corresponding to 450 absorbed photons per pixel. This is due  to the high amplifier gain used (ISO 3200). When saturation occurs, the Fano factor decreases, as the output is a constant.
At low illumination intensities, we measure a Fano factor much greater than 1, due to detector technical noise.

Image sensors such as CCD and CMOS have various sources of noise: thermal noise, leakage current and readout noise. Thermal and leakage noise accumulate with integration time, so that it is possible to eliminate them using short exposure times (of the order of a millisecond). In this case, readout noise becomes the dominant technical noise, and is given by the readout circuit, the amplifier and the analog to digital converter (ADC).
In image sensors, noise is usually counted in electrons (\SI{}{e^-}). The CCD camera and CMOS camera have a noise of \SI{10}{e^-} and \SI{3.3}{e^-} respectively.

\subsection{Random number generation}
To generate random numbers we illuminate the cameras so that the mean number of absorbed photons $\bar n$ is sufficient to give a quantum uncertainty $\sigma_q=\sqrt{\bar n}$ as large as possible whilst not saturating the detectors. In practice we illuminate the ATIK and Nokia cameras with an amount of light sufficient to generate \SI{1.5e4}{e^-} and \SI{410}{e^-} respectively.

Using equation~\ref{eq:entropy}, it is possible to calculate that the amount of entropy of quantum origin per pixel is 8.9 bits and 6.4 bits for each camera respectively, which are encoded over 16 and 10 bits, resulting in an average entropy per output bit of 0.56 for the CCD and 0.64 for the CMOS sensor.
Working parameters and results are summarised in table~\ref{table_results}.
\begin{figure}[htbp]
\begin{center}
\includegraphics[width=0.8\columnwidth]{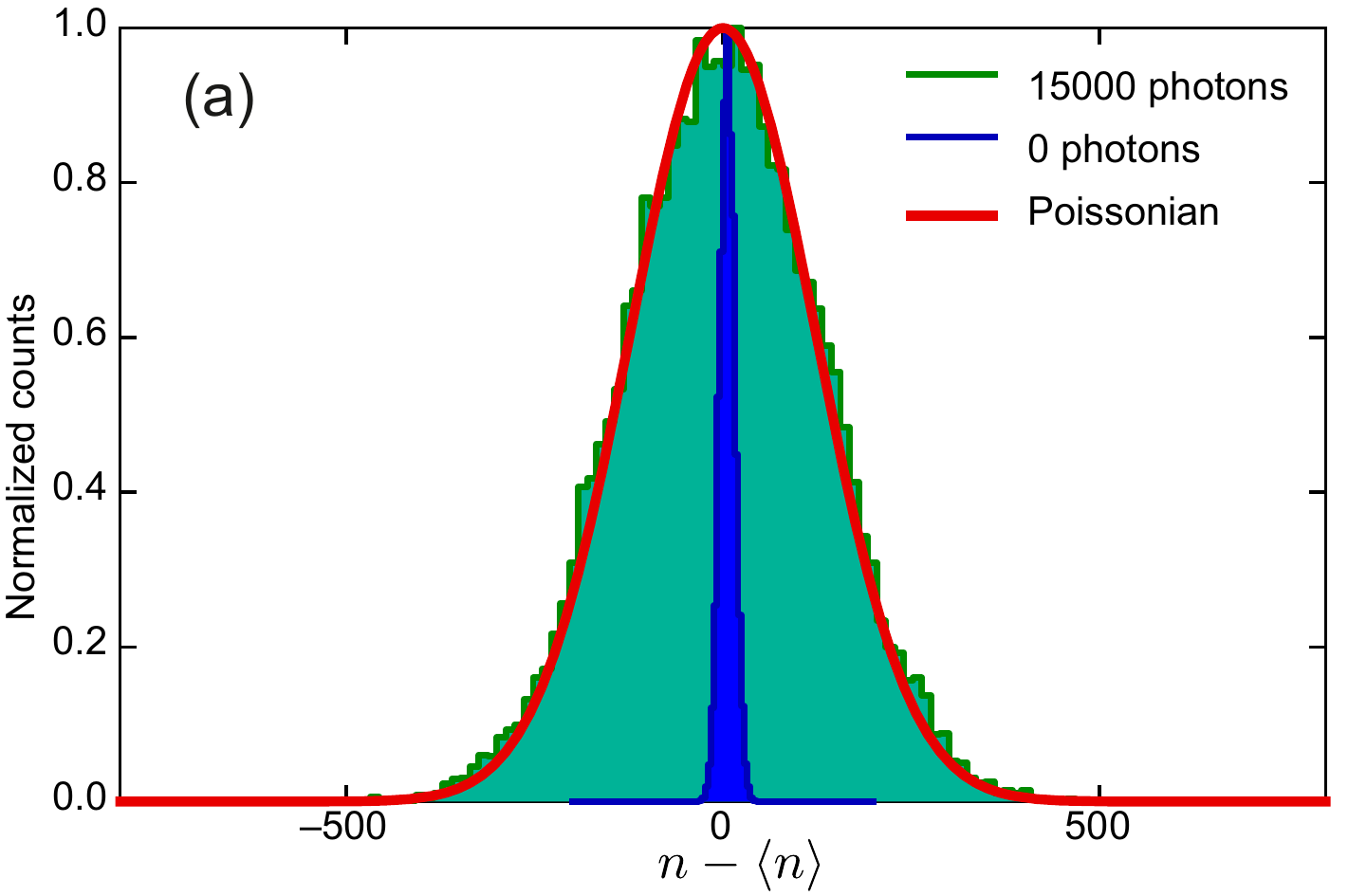}
\includegraphics[width=0.8\columnwidth]{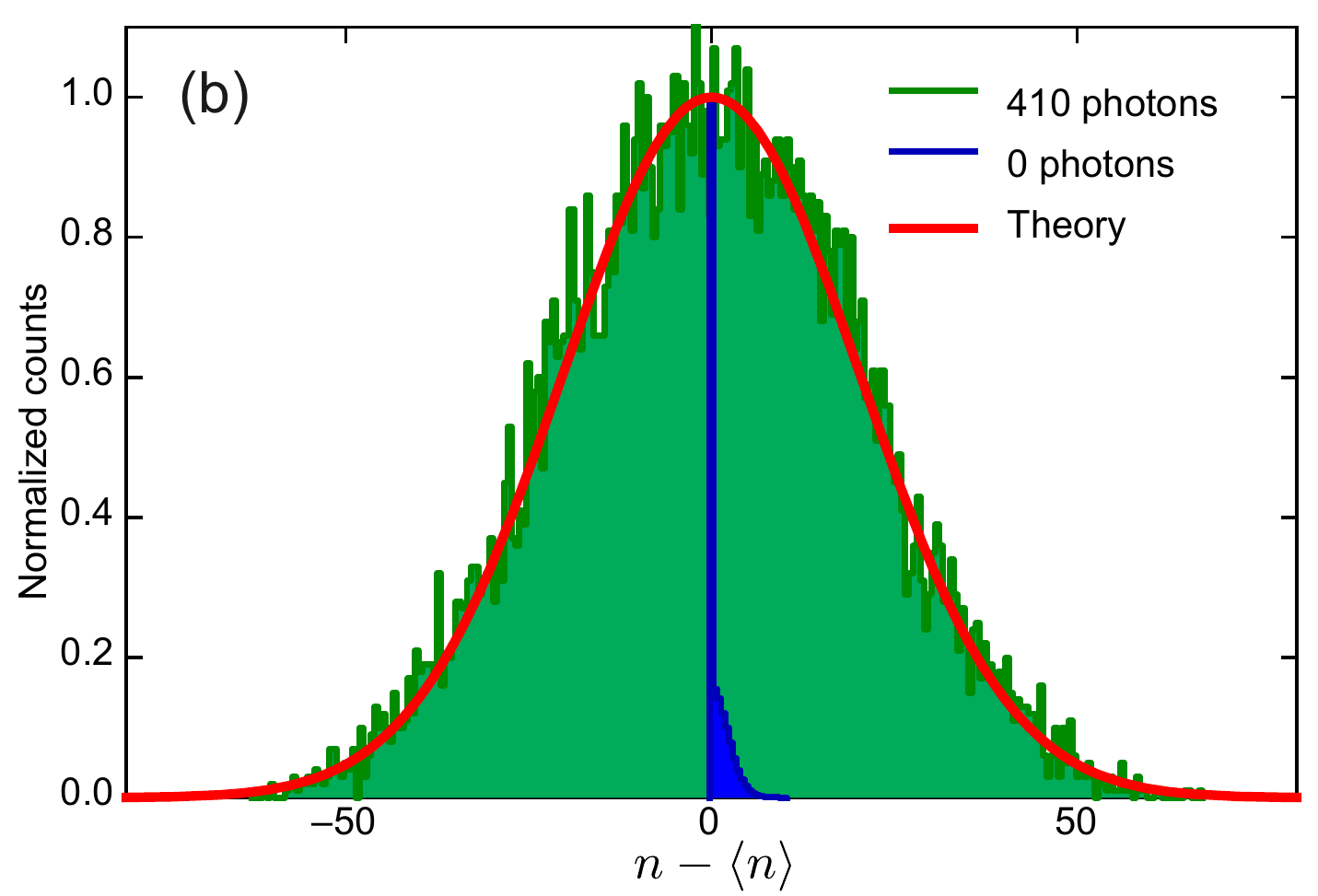}
\caption{Measurement of the quantum and classical noise of our ATIK (a) and Nokia (b) detectors. At the operating conditions quantum noise strongly dominates.}
\label{fig:ATIK_results}
\end{center}
\end{figure}

\begin{table}
	\begin{center}
		\begin{tabular}{l||c|c}
		\hline
			 & ATIK 383L & Nokia N9 \\ \hline
			Noise, $\sigma_t$ ($e^-$) & 10 & 3.3 \\
			Saturation ($e^-$) & \SI{2e4}{}  & \SI{500}{}  \\
			Illumination ($e^-$) & \SI{1.5e4}{} & \SI{410}{} \\
			Quantum uncertainty, $\sigma_q$ ($e^-$) & \SI{122}{} & \SI{20}{} \\
			Offset ($e^-$) & \SI{144}{} & \SI{-6}{} \\
			Output bits per pixel & 16 & 10 \\
			Quantum entropy per pixel & 8.9 bits & 6.4 bits \\
			Quantum entropy per raw bit & 0.56 & 0.64
		\end{tabular}
		\caption{Experimental parameters for the two cameras employed in this experiment.}
		\label{table_results}
	\end{center}
\end{table}

From the equation above, we calculate that using the camera in the Nokia cell phone, and an extractor with a compression factor of 4, for example with $k=500$ and $l=2000$, it would take an impossible $\sim\SI{e118}{}$ trials to notice a deviation from a perfectly random bit string. If everybody on earth used such a device constantly at 1Gbps, it would take \SI{e80}{} times the age of the universe for one to notice a deviation from a perfectly random bit string.


\section{Results and tests}
We collected 48 frames corresponding to approximately 5 Gbits of raw random numbers and processed them on a computer through an extractor with a 2000 bit input vector and a 500 bit output vector to generate 1.25 Gbits of random numbers. Random number generators are notoriously hard to test, however it is possible to check the generated bit string for specific weaknesses. The first step is to individuate potential problems of the system, and then test for them. First, we tested the generated random bit string before extraction. At this stage, the entropy per bit is still considerably less than unity; moreover, possible errors could arise from dead pixels and from correlations between pixels values given by electrical noise.

Besides increasing the mean entropy per bit, the randomness extractor also ensures that if some of the pixels become damaged, covered by dust or suffer from any other problem, an extremely good quality of the randomness is maintained.


Finally, we performed the ``die harder'' battery of randomness tests on both the extracted bit strings. This set of tests contains the NIST test, the diehard tests and some extra tests. The RNG passed all tests.

\subsection{Speed}
For many applications, such as the generation of cryptographic keys or gaming, speed is not as important as the affordability and portability given by this system. Nevertheless, a quantum random number generator based on an image sensor can provide very reasonable performance in terms of speed.
Consumer grade devices acquire data at rates between 100 Megapixels per second and 1 Gigapixel per second. After the necessary processing, each pixel will typically provide 3 random bits so that rates between 300 Mbps and 3 Gbps can be obtained. To sustain such high data rates, processing can be done either on an Field Programmable Gate Array (FPGA), or could be embedded directly on a CMOS sensor chip. Implementing the extractor fully in the software of a consumer device can sustain random bit rates greater than 1 Mbps, largely sufficient for most consumer applications.

\section{Conclusion and outlook}
We demonstrate a generator of random numbers of quantum origin using technology compatible with consumer and portable electronics. We believe that the simplicity and performance of this device will make the widespread use of quantum random numbers a reality, with an important impact on information security.

%
\end{document}